\documentclass[prd]{revtex4}
\usepackage{graphicx}
\usepackage{dcolumn}
\usepackage{bm}
\begin{document}

\title{Numerical simulation of a possible counterexample to cosmic
censorship}

\author{David Garfinkle}
\email{david@physics.uoguelph.ca}
\affiliation{Department of Physics, University of Guelph, Guelph, Ontario,
Canada N1G 2W1
\\ and Perimeter Institute for Theoretical Physics, 
35 King Street North, Waterloo Ontario, Canada N2J 2W9}

\begin{abstract}
A numerical simulation is presented here of the evolution of initial data 
of the kind that was conjectured 
by Hertog, Horowitz and Maeda \cite{Gary} to be a violation 
of cosmic censorship.
That initial data is essentially a thick domain wall connecting two regions
of anti-deSitter space.  The initial data has a free parameter that is the 
initial size of the wall.  
The simulation shows no violation of cosmic censorship, 
but rather the 
formation of a small black hole.  The simulation described here is for
a moderate wall size and leaves open the possibility that 
cosmic censorship might be violated for larger walls. 

\end{abstract}
\maketitle

\section{Introduction}
 
One of the outstanding questions in general relativity is that of cosmic 
censorship: the question of whether the singularities formed in gravitational
collapse are hidden inside black holes.  One thing that makes this issue 
difficult is that it is not clear what we might expect to be true.  It is
fairly straightforward to find spacetimes with naked singularities; however,
it can certainly be argued that those spacetimes are not generic (as in the 
case of critical collapse\cite{Matt}) or that the matter is not ``physically
reasonable'' (as in the case of dust).
The question is what is to be meant
by ``physically reasonable'' and whether cosmic censorship holds if the
condition of physically reasonable matter (as well as genericity) is imposed.
One might think that pointwise energy conditions would be a condition that 
would be satisfied by physically reasonable matter.  However, 
Hertog, Horowitz and Maeda\cite{Gary}
point out that matter can violate pointwise energy conditions and 
still give rise to a positive mass theorem.  They argue that such matter
should be considered physically reasonable; but that it gives rise to naked
singularities.  In reference \cite{Gary} initial data is 
generated and it is argued that
the evolution of that data gives rise to a singularity that cannot be hidden
inside a black hole.  
Such a singularity would have to either (i) be visible to an observer
at infinity or (ii) itself extend to infinity.  In reference\cite{Gary}
it is argued that possibility (i) is more likely; however, a recent
theorem of Dafermos\cite{dafermos} rules out possibility (i).  Thus
if the arguments of reference\cite{Gary} are correct then their initial
data evolves to a singularity that extends out to infinity.
The system is gravity coupled to a scalar field with a
potential that has two minima below zero.  The initial data is essentially
a thick domain wall that interpolates between the true vacuum on the inside
and the false vacuum on the outside.   
The arguments of reference\cite{Gary} are somewhat
heuristic, so it is worthwhile to evolve their initial data to see whether 
a naked singularity does indeed form.  
Alcubierre {\it et al} \cite{Miguel} perform numerical simulations
to evolve an initial data set for a system 
that has some similarities to the 
system considered in reference\cite{Gary}.  
Here the system is also gravity coupled
to a scalar field with a potential with two minima; and the initial
data is a thick domain wall that interpolates between the two vacuua.
However, in this case
one minimum of the potential is at zero and it seems unlikely that the 
positive mass theorem holds for this system.  The results of \cite{Miguel}
are that the domain wall accelerates outward, approaching the speed of light.

Since the results to be obtained may depend crucially on the type of matter 
used, it is important to simulate the system described in
reference\cite{Gary}
rather than a different system that may or may not be analogous.  In this
work we perform a preliminary version of such simulations.  
The work is preliminary for the following reason: the argument of
reference\cite{Gary} says that the singularity cannot be hidden inside
a horizon if the initial wall is sufficiently large.  Here, sufficiently
large means a radius greater than about 600 in units of the anti-deSitter
radius of curvature.  Simulating such a large wall is numerically very
challenging.  Instead, this simulation is for a wall of the more moderate
initial size of about 7.
The system and data, as well as the 
numerical methods are described in section II.  Results are presented in 
section III and conclusions in section IV.  

\section{Equations and numerical methods}

The system to be studied is a spherically symmetric scalar field $\phi$
with a potential $V$.  The appropriate equations are therefore the
Einstein-scalar equations:
\begin{eqnarray}
{G_{ab}} &=&  {\nabla _a} \phi {\nabla _b} \phi - {g_{ab}}
( {\textstyle {1 \over 2}} {\nabla ^c}\phi {\nabla _c} \phi + V )  
\label{einstein}
\\
{\nabla _a}{\nabla ^a} \phi &=& {{\partial V}\over {\partial \phi}}
\label{wave}
\end{eqnarray}
(Here we are using units where $8 \pi G =1$).
We use polar-radial coordinates for the metric which puts it in the form
\begin{equation}
d {s^2} = - {\alpha ^2} d {t^2} + {a^2} d {r^2} + {r^2} \left ( d 
{\theta ^2} + {\sin ^2} \theta d {\varphi ^2} \right )
\label{metric}
\end{equation}
It is helpful to define the quantities $X$ and $Y$ given by
\begin{eqnarray}
X &\equiv & {{\partial \phi}\over {\partial r}}
\\
Y &\equiv & {a \over \alpha} {{\partial \phi}\over {\partial t}}
\end{eqnarray}
from which it follows that equation of motion for $\phi$ is
\begin{equation}
{{\partial \phi} \over {\partial t}} = {\alpha \over a} Y
\label{dtphi}
\end{equation}
From equation (\ref{wave}) it follows that the equation of motion for $Y$ is
\begin{equation}
{{\partial Y} \over {\partial t}} = {1 \over {r^2}} {\partial \over 
{\partial r}} \left ( {r^2} {\alpha \over a} X \right ) - \alpha a 
{{\partial V} \over {\partial \phi}}
\label{dty}
\end{equation}
From equation (\ref{einstein}) we obtain the following constraint equations
for $a$ and $\alpha$
\begin{eqnarray}
{{\partial a}\over {\partial r}} &=& {{a(1-{a^2})}\over {2 r}} + 
{\textstyle {1 \over 4}} r a \left ( {X^2} + {Y^2} + 2 {a^2} V \right )
\label{dra}
\\
{\partial \over {\partial r}} \ln (a\alpha ) &=& {r \over 2} \left ( 
{X^2} + {Y^2}\right ) 
\label{draalpha}
\end{eqnarray}
We now turn to numerical methods.  The equations simulated are equations
(\ref{dtphi}-\ref{draalpha}).  We use unequally spaced values of $r$.    
At grid point $i$ the quantity $X$ is calculated as 
${X_i} = ({\phi _{i+1}}-{\phi_{i-1}})/({r_{i+1}}-{r_{i-1}})$.  
For any quantity $f$ define quantities $f_+$ and $f_-$ by 
${f_+} = (f_{i+1}+{f_i})/2$
and ${f_-}=({f_i}+{f_{i-1}})/2$.  Also define the quantities $X_p$ and
$X_m$ by 
${X_p} = ({\phi_{i+1}}-{\phi _i})/({r_{i+1}}-{r_i})$
and ${X_m}=({\phi _i}-{\phi_{i-1}})/({r_i}-{r_{i-1}})$.   
The first term on the right hand side of equation (\ref{dty}) is 
differenced as
\begin{equation}
{1 \over {r^2}} {\partial \over {\partial r}} \left ( {r^2}
{\alpha \over a} X \right ) \to {3 \over {{r_+ ^3}-{r_- ^3}}} \left ( 
{r_+ ^2} {{\alpha _+}\over {a_+}} {X_p} - {r_- ^2} {{\alpha _-}\over {a_-}}
{X_m}\right )
\end{equation} 
Equations (\ref{dtphi}-\ref{dty}) are
integrated using a three step iterated Crank-Nicholson method\cite{ICN}
with Kreiss-Oliger dissipation\cite{KO} used in equation (\ref{dty}).
Each iteration of the Crank-Nicholson method involves an integration
of equations (\ref{dra}-\ref{draalpha}) which is done using a second
order predictor-corrector method. 

At the origin, (gridpoint $i=1$) smoothness requires that $\phi$ and
$Y$ have vanishing derivative with respect to $r$.  We impose this as
${\phi_1}=(4{\phi_2}-{\phi_3})/3$ and correspondingly for $Y$.  We also
impose the condition ${a_1}=1$ which is required by smoothness of the
metric.  The value of $\alpha$ at the origin can be freely specified, and
we choose ${\alpha _1}=1$.  At large distances, the spacetime should be in
the anti-deSitter false vacuum corresponding to $\phi={\phi _{\rm vac}}$.
We apply this condition by  imposing ${\phi _N} = {\phi _{\rm vac}}$ and
${Y_N}=0$ where $N$ is the last gridpoint.  

Stability of the simulations requires that the time step satisfy a 
Courant condition.  We choose $dt$ to be the minimum over all gridpoints
$i$ of ${1\over2} (r_{i+1}-{r_i}){a_i}/{\alpha _i}$.  This Courant condition
is one reason for the use of unequally spaced $r_i$.  The presence of vacuum
energy leads to a large variation in $a/\alpha $ which would then lead to 
an extremely small time step if we had a uniform spacing for $r$. 

We now turn to a consideration of the initial data.  The potential used in 
reference\cite{Gary} is 
\begin{equation}
V(\phi ) = -3 + 50 {\phi ^2} - 81 {\phi ^3} + k {\phi ^6}
\end{equation}
where $k$ is a constant whose value will be specified later.   The false
vacuum is at $\phi =0$ while the true vacuum is at some other value
$\phi = {\phi _{\rm vac}}$.   The approach of reference\cite{Gary} to choosing 
initial data is the following: consider initially
static field configurations that are in the false vacuum at 
$r=0$ and in the true vacuum for 
$ r>{R_{\rm vac}}$ for some constant $R_{\rm vac}$.  
Then the contribution of the potential to
the total mass is 
\begin{equation}
{m_V}= {\textstyle {1 \over 2}} {R_{\rm vac} ^3} {\int _0 ^1} 
{e^{- {\int _y ^1} d {\hat y} {\hat y} {{\phi '}^2}/2}} V {y^2} d y
\end{equation}
Here $y=r/{R_{\rm vac}}$ and ${\phi '}=\partial \phi /\partial y$.  
Define $\rho _V$ to be the minimum over all field configurations  of
$2 {m_V} {R_{\rm vac} ^{-3}}$ and choose as initial data the field
configuration for which the minimum is attained.   
The positive energy theorem will hold provided that ${\rho _V} >
V({\phi _{\rm vac}})/3$.  For the present potential, this inequality 
is just barely satisfied for $k=45.928$.  Define $x = \ln y$.  Then 
minimizing $\rho _V$ corresponds to the differential equation
\begin{equation} 
{{{d^3} \phi}\over {d {x^3}}} = {{{d^2} \phi}\over {d {x^2}}} \left [
3 + {\textstyle {1 \over 2}} {{\left ( {{d\phi}\over {dx}}\right ) }^2}
+ {{2 V {{{d^2} \phi}\over {d {x^2}}} + {{dV}\over {d\phi}}
{{\left ( {{d\phi}\over {dx}}\right ) }^2} + {{{d^2}V}\over {d {\phi^2}}}
{{d\phi}\over {dx}}}\over {V {{d\phi}\over {dx}}+ {{dV}\over {d\phi}}}}
\right ]
\label{initode}
\end{equation} 
We find the solution to this equation using a shooting method.  Smoothness
and the condition that the field be in the false vacuum at $r=0$ yields the 
condition that for large negative $x$ we have $\phi = c{e^{\beta x}}  $ 
where $c$ is a constant and $\beta =({\sqrt {409}} -3)/2 \approx 8.6$. 
We integrate equation (\ref{initode}) from large negative $x$ to $x=0$
using the fourth order Runge-Kutta method.  The value of $c$ is found
using a binary search to be the one that yields $\phi = {\phi _{\rm vac}}$
at $x=0$.   

The spacing in $r$ is chosen as follows: choose a constant $R_{\rm max}$ and
a coordinate $\tilde r$ equally spaced from zero to $R_{\rm max}$.  Then
define $r$ by
\begin{equation}
r={\textstyle {1\over 2}} \left [ 1.1 {\tilde r} + 0.9 \ln 
\left ( {{\cosh ({\tilde r} - {R_{\rm vac}})}\over {\cosh {R_{\rm vac}}}}
\right ) \right ]
\end{equation}

\section{Results}

\begin{figure}
\includegraphics[scale=0.8]{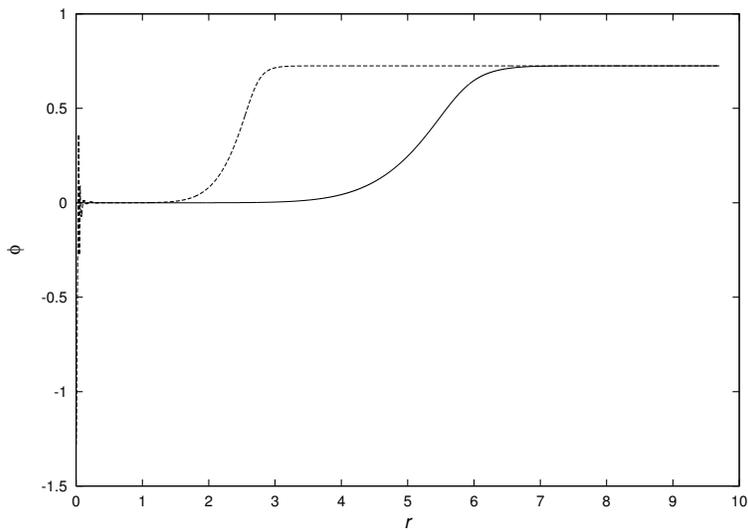}
\caption{\label{phis} the scalar field $\phi$ at the initial time
(solid line) and the final time (dotted line)}
\end{figure}

\begin{figure}
\includegraphics[scale=0.8]{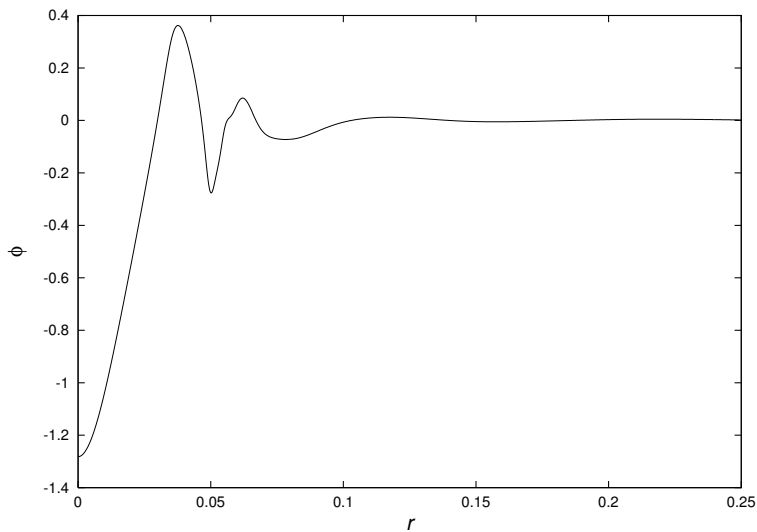}
\caption{\label{phifin} the scalar field $\phi$ at the final time}
\end{figure}

\begin{figure}
\includegraphics[scale=0.8]{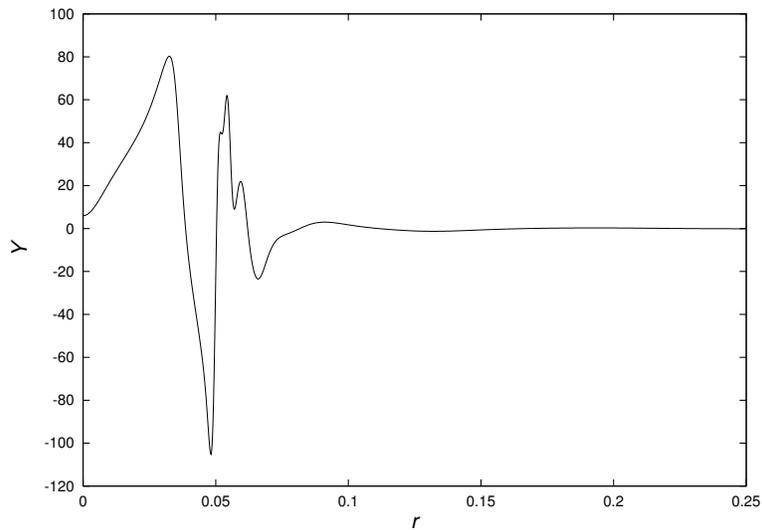}
\caption{\label{yfin} $Y$ at the final time}
\end{figure}

\begin{figure}
\includegraphics[scale=0.8]{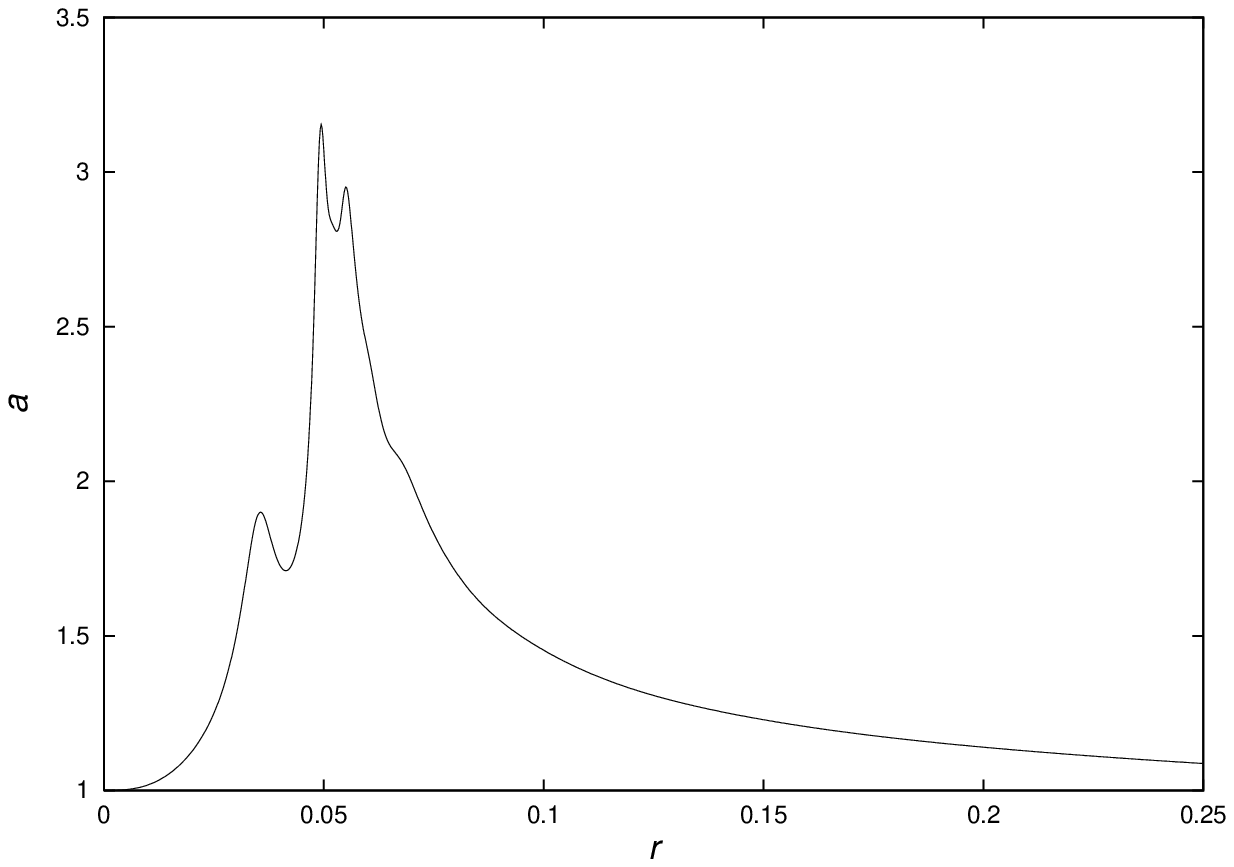}
\caption{\label{afin}  $a$ at the final time}
\end{figure}

The simulation was done in double precision on a SunBlade 2000.  
The number of gridpoints was 8001 and the values of the parameters were
as follows:
${R_{\rm vac}}=7 $ and $ {R_{\rm max}}=16$ while $k$ was chosen so that
${\phi _{\rm vac}}=0.725$ which yields the critical value of $k$. 
The coordinate system used (equation (\ref{metric})) cannot evolve
past the formation of a trapped surface, since it assumes that $r$
is spacelike, whereas $r$ becomes null on a trapped surface.  As a 
trapped surface forms, the metric quantity $a$ grows without bound.
The simulation is run until $a$ grows large enough to signal that a
trapped surface is forming.
Figure \ref{phis} shows the scalar field $\phi $
at the initial time (solid line) and at a time shortly before the 
formation of a trapped surface (dotted line).  This time is close to
$\pi /2$.  What happens in the evolution of
$\phi$ is that the initial domain wall moves inward somewhat; but in 
addition, a wave pulse comes off the wall, moves to the center and forms
a black hole there.  Since the most interesting features of the final 
configuration are on a small spatial scale, subsequent figures will plot
those features in the range $0<r<0.25$.  
Figures \ref{phifin}-\ref{afin} show respectively $\phi , \; Y$ and $a$
at the time shortly before the formation of a trapped surface.

Note from figure \ref{yfin} that the time derivative of the scalar field is
getting quite large at the final time and at points close to the center.
Thus a region of high curvature is forming near the center.
Figure \ref{afin} indicates the formation of a trapped surface at
$r \approx 0.05$.
What is important here is not simply the formation of a trapped surface, but
also that the size of the trapped surface is of the same order as the size
of the region of high curvature rather than much smaller than the region of
high curvature.  This makes it likely that what is happening is the formation
of a black hole that will contain the singularity.  
Note that the size of the black hole is considerably smaller than the spatial
scale set by the initial size of the wall.

\section{Conclusions}

The results of this simulation of a moderate size wall are quite 
different from what is conjectured in reference\cite{Gary} to
be the behavior of a large wall.
Instead of a naked singularity
one gets a small black hole.  Instead of the collapse of a rather large 
homegeneous region, one has very localized and decidely non-homogeneous 
behavior near the center of spherical symmetry.  The reason for this 
behavior follows from properties of the initial data: note that near
the center the scalar field is proportional to $r^\beta$ where $\beta 
\approx 8.6$.  This large power means that the scalar field is essentially
zero in a spherical region around the center.  In contrast, the analysis
of reference\cite{Gary} assumes that in the central region 
the scalar field can
be modeled as homogeneous but time dependent.   Since the initial scalar
field in the central region is negligible, subsequent behavior of the
scalar field in the central region must come from modes that propagate 
inwards, in this case from the domain wall.  As these modes propagate
inwards, their amplitude grows and their wavelength shortens until finally
they form a small black hole at the center.

Thus the results of this simulation do not provide a counterexample
to cosmic censorship.
Note, however that this 
simulation is done with a moderate value of $R_{\rm vac}$ 
(${R_{\rm vac}}=7$) while the argument of reference \cite{Gary} claims that 
violations of cosmic censorship occur for sufficiently large $R_{\rm vac}$
(${R_{\rm vac}}>600$).
Thus it is still possible that this model violates cosmic censorship but
not for moderate values of $R_{\rm vac}$ like the one used in this simulation.
Put another way, the arguments of reference\cite{Gary} do not lead one to
expect the formation of a naked singularity from the ${R_{\rm vac}}=7$ 
form of their initial data.  Therefore the fact that this simulation finds
black hole formation does not contradict the arguments of reference\cite{Gary}.
Nonetheless, the details of the collapse process tend to cast doubt on 
some parts of the argument of reference\cite{Gary}.  In particular, 
regardless of the size of the wall, the initial data for the scalar field 
behaves like $r^\beta$ near the center for $\beta \approx 8.6$.  Thus 
the initial data in the central region does not look like anti-de Sitter
space with a small homogeneous perturbation.  However, the argument 
of reference\cite{Gary} depends on treating the evolution of the
central region as the evolution of a homogeneous space.  Thus there is
reason to doubt this part of the argument.  To resolve this issue
one would need to do a simulation with large $R_{\rm vac}$.
Since in anti-de Sitter space $\alpha $ and $a^{-1}$ grow linearly
with $r$, large values of $R_{\rm vac}$ involve large values of $\alpha$
and $a^{-1}$.  It is therefore likely that the numerical method used in 
this work is not sufficiently robust to treat large values of $R_{\rm vac}$
and that some other numerical method will have to be used.  This problem
is under study.

Finally, note that the authors of reference \cite{Gary} have proposed another
model\cite{Gary2} that they claim leads to violations of cosmic censorship.
However, numerical simulations of that model done by Gutperle and Kraus 
\cite{Kraus} indicate that that model leads to the formation of black
holes rather than naked singularities.

\section{Acknowledgements}

I would like to thank Gary Horowitz, Rob Myers, Eric Poisson and Mihalis
Dafermos for helpful discussions.
This work was partially supported by NSF grant PHY-0244683 to Oakland 
University.

\end{document}